\pdfoutput=1
\documentclass[12pt]{article}

\usepackage[a4paper,
            mag=1000, includefoot,
            left=3cm, right=1.5cm, top=2cm, bottom=2cm, headsep=1cm, footskip=1cm]{geometry}
\usepackage[english]{babel}

\usepackage{graphicx}        
\usepackage{latexsym,amssymb}
\usepackage{amsmath}
\usepackage{bm}
\usepackage[shortlabels]{enumitem}
\usepackage{algpseudocode}
\usepackage{url}
\usepackage{fancyvrb}
\usepackage{xspace}
\usepackage{hyperref}
\hypersetup{ colorlinks = true, citecolor = blue, urlcolor = blue}

\graphicspath{{img/}}

\usepackage{euscript}

\newcommand{\cT}{\EuScript{T}}

\newcommand{\rmT}{\mathrm{T}}

\newcommand{\tX}{\mathsf{X}}

\newcommand{\bfX}{\mathbf{X}}

\newcommand{\bt}{\begin{theorem}}
\newcommand{\et}{\end{theorem}}
\newcommand{\bl}{\begin{lemma}}
\newcommand{\el}{\end{lemma}}
\newcommand{\bp}{\begin{proposition}}
\newcommand{\ep}{\end{proposition}}
\newcommand{\bc}{\begin{corollary}}
\newcommand{\ec}{\end{corollary}}

\newcommand{\bd}{\begin{definition}\rm}
\newcommand{\ed}{\end{definition}}
\newcommand{\bex}{\begin{example}\rm}
\newcommand{\eex}{\end{example}}
\newcommand{\br}{\begin{remark}\rm}
\newcommand{\er}{\end{remark}}

\newcommand{\btbh}{\begin{table}[!htbp]}
\newcommand{\etb}{\end{table}}
\newcommand{\bfgh}{\begin{figure}[!htbp]}
\newcommand{\efg}{\end{figure}}

\newcommand{\bea}{\begin{eqnarray*}}
\newcommand{\eea}{\end{eqnarray*}}
\newcommand{\be}{\begin{eqnarray}}
\newcommand{\ee}{\end{eqnarray}}

\newtheorem{proposition}{Proposition}
\newtheorem{corollary}{Corollary}
\newtheorem{theorem}{Theorem}
\newtheorem{remark}{Remark}
\newtheorem{lemma}{Lemma}
\newtheorem{definition}{Definition}

\newcommand{\projVec}{W}

\newlist{arguments}{description}{1}
\setlist[arguments]{style=sameline}
\newlist{inlinelist}{enumerate*}{1}
\setlist*[inlinelist,1]{label=(\alph*), itemjoin={{, }}, itemjoin*={{, and }}}

\algnewcommand\algorithmicinput{\textit{Input:}}
\algnewcommand\Input{\item[\algorithmicinput]}
\algnewcommand\algorithmicoutput{\textit{Output:}}
\algnewcommand\Output{\item[\algorithmicoutput]}
\newcommand{\algrule}[1][1pt]{\par\vskip.5\baselineskip\hrule height #1\par\vskip.5\baselineskip}

\newcounter{ssaalg}[section]
\makeatletter
\renewcommand\thessaalg{\thechapter.\@arabic\c@ssaalg}%
\def\ext@ssaalg{loa}
\def\fnum@ssaalg{\textsc{Algorithm~\thessaalg}}
\makeatother

\DefineVerbatimEnvironment{Code}{Verbatim}{fontsize=\small,fontshape=sl}
\DefineVerbatimEnvironment{Func}{Verbatim}{fontsize=\small,fontshape=tt}
\DefineVerbatimEnvironment{CodeInput}{Verbatim}{fontsize=\footnotesize, fontshape=sl}
\DefineVerbatimEnvironment{CodeInput}{Verbatim}{fontsize=\scriptsize, fontshape=sl}

\makeatletter
\newcommand\code{\bgroup\@makeother\_\@makeother\~\@makeother\$\@codex}
\def\@codex#1{{\small\ttfamily\hyphenchar\font=-1 #1}\egroup}
\makeatother
\newcommand{\pkg}[1]{\textsc{#1}}
\let\proglang=\textsf
\def\R{{\normalfont\ttfamily R}\xspace}

\makeatletter
\def\verbatim@font{\small\ttfamily}
\makeatother

\def\spaceR{\mathbb{R}}

\def\sspan{\mathop{\mathrm{span}}}

\makeatletter
\def\adots{\mathinner{\mkern2mu\raise\p@\hbox{.}
\mkern2mu\raise4\p@\hbox{.}\mkern1mu
\raise7\p@\vbox{\kern7\p@\hbox{.}}\mkern1mu}}
\newcommand{\l@abcd}[2]{\hbox to\textwidth{#1\dotfill #2}}
\makeatother

\begin{document}

\author{Nina Golyandina\footnote{St.Petersburg State University, Universitetskaya nab. 7/9, St.Petersburg, Russia. n.golyandina@spbu.ru}\footnote{The reported study was funded by RFBR, project number 20-01-00067.}}
\title{Detection of signals by Monte Carlo singular spectrum analysis: Multiple testing}
\date{}
\maketitle

\begin{abstract}  Detection of a signal in a noisy time series using Monte Carlo singular spectrum analysis (MC-SSA) is studied from the statistical viewpoint. The MC-SSA test consists of simultaneous testing of several hypotheses related to the presence of different frequencies. The multiple MC-SSA test procedure is constructed to control the family-wise error rate. The technique to control both the type I and the type II errors and also
to compare criteria is proposed to study several versions of MC-SSA.
\end{abstract}


\section{Introduction}

Singular spectrum analysis (SSA) solves a wide range of problems of time series analysis and image processing (see \cite{Golyandina.etal2018} for examples and references).
Here we consider a specific problem of detection of a signal (e.g., a periodic component) in a noisy time series.

The common scheme of the decomposition stage of SSA consists of the construction of a so-called trajectory matrix $\bfX$
from the initial object $\tX$ and an expansion of this trajectory matrix into a sum of
elementary components, which are ordered by their contribution. In the basic version of SSA, this expansion
is performed using the singular value decomposition (SVD). The left singular vectors of the SVD are eigenvectors of the matrix $\bfX\bfX^\rmT$, which is an estimate of the autocovariance matrix of the time series if its mean equals zero or the time series is centred. The reconstruction stage starts with the identification of elementary components corresponding to a component under interest (e.g. a signal)
and then the reconstruction of this time series component through a grouping of the identified components is performed.

Generally, SSA is a model-free method. However, a considerable part of the SSA theory is devoted to
the extraction of time series components, which are governed (maybe, approximately)
by a linear recurrence relation \cite{Golyandina.etal2018}; in particular, a sum of sine waves with slowly-varying amplitudes belong to this class of signals.
SSA extracts periodic components well; however, it is well-known that
the extracted components can be spurious, since they can be produced by noise. In a sense, this is a payment for
the nonparametric nature of the method. If we want to apply the statistical approach for judging, a model should be assumed.
 Usually, the question about the existence of a signal in the time series is formulated as testing the hypothesis that the series is a stochastic process. The criterion should be powerful against the alternative hypothesis, which states the existence of a non-random (e.g. periodic) component. There are a lot of statistical criteria for testing these hypotheses for different classes of stochastic processes (see e.g. \cite{Vaughan2005} for red noise or \cite{Louca2015} for an Ornstein-Uhlenbeck state space, which is a continuous  analogue of the first-order autoregressive processes AR(1)).
We consider the construction of such a criterion in the framework of SSA, since SSA is able to reconstruct the detected time series component.

It follows from the properties of SSA (see e.g. the description of the relation between the spectral density and the eigenvectors in \cite[Sect. 6.4.3]{Golyandina.etal2001}, where the results from \cite{GrS58} in terms of SSA are discussed) that a natural assumption for the noise model is that noise is red (the
AR(1) with a positive coefficient). This is because the spectral density of red noise is monotonic;
therefore, the eigenvectors are similar to sinusoids and can be connected to different frequencies. Recall that $\bm{\xi}=(\xi_1,\ldots,\xi_{N})$ is red noise with parameters $\varphi$ and $\delta$ if $\xi_n = \varphi \xi_{n-1} + \delta \epsilon_n$, where $0 < \varphi < 1$, $\epsilon_n$ is white Gaussian noise with mean 0 and variance 1 and
$\xi_1$ has a normal distribution with mean zero and variance $\delta^2/(1-\varphi^2)$.
Moreover, in climatology, the common model is a weak signal (if any) in red noise.
Both considerations inspired the creation of the Monte Carlo SSA method (MC-SSA).
MC-SSA was proposed in \cite{Allen.Smith1996} and later was considered in many papers
(\cite{Allen.Robertson1996}, \cite{Garnot.etal2018}, \cite{Greco.etal2011}, \cite{Jemwa.Aldrich2006}, \cite{Palus.Novotna1998}, \cite{PALUS.Novotna2004}, among others).

However, the terminology in these papers differs from the standard statistical terminology
and therefore it is very important to bridge the gap between the applied and statistical approaches.
Moreover, this can help to avoid wrong conclusions in real-life applications.
Investigation of criterion properties (and comparison with other tests)
will be performed by the estimation of the type I and type II errors.

\smallskip
Let us briefly described how the basic MC-SSA test is constructed.
At Decomposition step of SSA,
each decomposition component (related to the eigenvectors of the sample autocovariance matrix, in the basic version) can be put into correspondence with a frequency, since the eigenvectors of the autocovariance matrix of red noise, which is a Kac–Murdock–Szeg\H{o} Toeplitz matrix, are sine waves  with almost equidistant frequencies \cite{Trench2010}. The eigenvalues are equal to the total squared norm of projection of the trajectory matrix columns on the eigenvectors and therefore reflect contributions of the decomposition components (and therefore of the corresponding frequencies) into the time series.
The original idea of MC-SSA is to estimate the parameters of red noise and apply the bootstrap simulations
 to construct prediction intervals for eigenvalues
in the case when there is no signal.
If an eigenvalue of the time series is beyond the constructed prediction intervals,
the corresponding eigenvector frequency is considered significant.
Moreover, it is possible to reconstruct the detected signal.
In the modifications of MC-SSA, the choice of vectors for projection can vary.

It is clear from the method description
that  there is the problem of multiple testing when the probability of the false detection of
a periodic component for one of the considered frequencies (family-wise error rate) is unknown and is much larger than
the given significance level (single-test error rate).
This problem is discussed in different papers devoted to MC-SSA.
The theoretical approach to multiple testing, which we propose in this paper,
allows constructing the multivariate criterion with the given family-wise error rate.

\textbf{Novelty.} The novelty of the paper is applying the statistical approach to the signal-detection
problem in the framework of Monte Carlo SSA to control the type I error
and estimate the type II error.
For simultaneous testing of the presence of multiple frequencies, a multiple version of MC-SSA is proposed to
control the family-wise error rate.
Basing on the numerical study, we discuss pluses and minuses of several versions of MC-SSA.

\textbf{Structure.} In Section~\ref{sec:stat}, we briefly describe the statistical approach to hypothesis testing.
 In Section~\ref{sec:MCSSA}, versions of the MC-SSA algorithms are presented, including the proposed multiple testing algorithms. Section~\ref{sec:study} is devoted to the investigation of the MC-SSA approach using the statistical approach.
All numerical studies were performed in \R{}, with the use of the \pkg{Rssa} package \cite{Rssa}.

\section{Statistical approach to hypothesis testing}
\label{sec:stat}

In papers starting from \cite{Allen.Smith1996}, the method MC-SSA is described
mostly as a method for applied problems and therefore the way of description
is not conventional for statisticians. Therefore, let us start with a statistical
approach to the problem.

Let the null hypothesis be that the time series is a pure stationary stochastic process.
In the considered context, it can be white or red noise. Sometimes, one says that the presence of a signal in noise is tested,
whereas the null hypothesis is formulated as the hypothesis about the absence of a signal in noise.
Consider a criterion, which determines if the null hypothesis is rejected or is not rejected.
If the null hypothesis is rejected at the given significance level $\alpha$, then one can claim the presence of a signal (more precisely, a deviation from the null hypothesis).
The probability to reject the null hypothesis if it is true is called type I error ($\alpha_I$).
If a criterion is correct, then the type I error is equal to the given significance level
(or at least not larger than $\alpha$).
Different criteria differ by the power against an alternative hypothesis. The power is
the probability to reject the null hypothesis if the alternative hypothesis is true.
The alternative hypothesis that the time series contains a periodic component is important
in practice; therefore, we will consider criteria, which are
powerful against such hypotheses.

If we want to make a false discovery rarely, then we choose a small significance level $\alpha$
and this guarantees that we will reject the  true null hypothesis with probability not larger than $\alpha$.
However, we can not choose a very small significance level, since the test power decreases as $\alpha$ decreases.

Note that it is not permitted to consider a criterion if its type I error exceeds the given significance level (a \emph{liberal} criterion).
Therefore, before a comparison of criteria by power, one should be sure that the type I error
lies in the given range.
If  the type I error is less than the significance level (a \emph{conservative} criterion), this is admissible; however,
this means that this criterion can be improved, that is, the power can be increased
by a correction to obtain the type I error closer to the significance level.

A useful characteristic of a criterion is the possibility to interpret
the difference from the null hypothesis if this hypothesis is rejected.

\subsection{Bootstrapping}
Most of the criteria have the following form. A constructed test statistic measures
the difference between data and the null hypothesis in some way. There is a threshold
such that if the test statistic is larger than the threshold, the null
hypothesis is rejected. Certainly, this threshold depends on the significance level
$\alpha$. It is not uncommon that this threshold can not be obtained theoretically.
Then simulations are used. Surrogate data are simulated according to the null hypothesis and the test statistic is calculated many times ($G$) to determine the threshold; this approach is widely used, see e.g. \cite{Lancaster2018}. Consequently, the threshold, for which the null hypothesis is rejected approximately $\alpha G$ times, is found. The estimated threshold is used for testing the hypothesis for real-life data. The surrogate data should be obtained exactly in the same way as the test statistic was generated for the original data.
The described approach can be called Monte-Carlo one. This approach helps to construct the criterion with
the type I error tending to $\alpha$ as $G$ tends to infinity.

However, the Monte Carlo approach can be applied if the null hypothesis fully determines the data.
For example, the null hypothesis states that the time series is red noise with variance
$\delta^2$ and coefficient $\varphi$, where $\delta^2$ and $\varphi$ are known numbers.

Unfortunately, this is not the case in practice. Therefore, the so-called bootstrapping is used
(``pull yourself up by your bootstraps''). If $\delta^2$ and $\varphi$ are unknown, then they
are estimated with the help of the real-life data under study and then the surrogate data
are produced by simulations with the estimated parameter.
Since the estimated parameters differ from the true (unknown) parameters,
then the  type I error can generally be far from $\alpha$ and the test can become liberal or conservative.
Thus, Monte-Carlo SSA (this is the name of a family of concrete algorithms) is a kind of testing with bootstrapping.

\subsection{Estimation of type I error and power}
The above considerations about the relation between the type I error, the significance level $\alpha$ and the
level of power can not be applied in practice, since the type I error and the power are unknown.
If something is unknown theoretically, then simulation helps again.

Let a criterion be constructed to make the decision (reject or not reject)
for a given significance level $\alpha$. It can be constructed theoretically or with
bootstrapping/simulations within, it does not matter.
Then the sample data with given parameters according to
the null hypothesis (this is the Monte-Carlo approach) are simulated many times ($M$). Then the proportion of
cases with the rejection of the null hypothesis is an estimate of the type I error.

To estimate power, we should set an alternative hypothesis.
There is freedom in the choice. The common rule is to include into
the alternative hypothesis the case, which is important, that is,
the case that should be distinguished from the null hypothesis.
For example, the alternative can state that the time series is a sine wave with a given frequency, amplitude and phase
corrupted by noise with the same parameters as were considered for the null hypothesis.

For power estimation, the procedure similar to that for the type-I error is fulfilled.
We simulate surrogate data with the given parameters according to
the alternative hypothesis many times ($M$). Then the proportion of
cases with the rejection of the null hypothesis is the estimate of the power.

\subsection{Prediction intervals for testing hypotheses}
Let a test statistic $t$ provide an interpreted characteristic of data (e.g. a contribution of a frequency to the observed time series).
Thus, the question about the presence of a signal can be reformulated as
``can this value of $t$ be caused by the noise component only?''.

The answer to the question can be obtained in the standard way. Since the contribution $t$ is random,
there is a prediction interval for it. The prediction interval can be constructed by simulation.
If we generated a sample of possible contributions, then the 95\% prediction interval is the interval between
the 2.5\% and 97.5\% quantiles of this sample. In statistical terminology, this interval is not called confidence, since confidence intervals are constructed for unknown (non-random) parameters and their length would tend to zero as the number of simulations (sample size) tends to infinity.

$\gamma$-Prediction intervals serve for testing hypotheses with a significance level $\alpha$ for
$\alpha = 1 - \gamma$. If the observed value of $t$ does not belong to the prediction interval, the null hypothesis
is rejected. It is convenient to depict, say, 95\%-prediction intervals for $t$ to visualize the hypothesis testing with the significance level 5\%.

\subsection{One-tailed and two-tailed criteria}
We mentioned that a criterion consists of a test statistic $t$ and a threshold $t_0$.
The use of this threshold can be different. Moreover, the threshold can consist of two numbers, $t_1$ and $t_2$.
For example, the null hypothesis can be
rejected if the test statistic is larger than $t_1$ or smaller than $t_2$ (two-tailed test) or
if $t>t_0$ (one-tailed test). The choice of the criterion type depends on the alternative hypothesis,
since we want to increase the power against the chosen alternative.

If we want to detect both cases, when the contribution of a frequency is either larger or smaller than that for  pure noise, then we choose a two-tailed test. If we want to detect only the excess of the frequency contribution, we choose a one-tailed test.
This approach can be expressed in terms of prediction intervals. If we are interested to find the frequency
with contribution larger than that of noise, then the one-sided prediction interval has the form $[0,t_0]$ (in the general case,
$[-\infty,t_0]$; however, in our case the test statistic is non-negative). In the two-tailed case, the two-sided prediction interval is $[t_1,t_2]$.

\subsection{Multiple testing}
The problem of multiple testing is well-known. The approach to the statistical testing described above is
applicable for single tests only, since the probability of false discovery is controlled for each individual test only.

If we test several tests ($m$) simultaneously, we are interested in the so-called family-wise error rate (FWER).
FWER is the probability of false discovery in at least one of $m$ tests.
This probability can be vastly larger than the chosen small $\alpha$.
Thus, we should not use a set of single tests with $\alpha$ if we want to control FWER. Ideally, we should construct
one multivariate test instead of several single tests. If this is technically hard, the Bonferroni
correction is used (performing single tests with significance level $\alpha/m$); this trick controls the FWER not larger
than $\alpha$ (usually, this does the family-wise type I error considerably smaller than $\alpha$; that is, the multiple testing is conservative and therefore decreases the test power). If the single tests are independent, the \v{S}id\'{a}k correction (performing single tests with significance level $1 - (1-\alpha)^{1/m}$) provides the exact test; however, in the general case, testing with the \v{S}id\'{a}k correction can be liberal and therefore the \v{S}id\'{a}k correction is not applied.

\section{Monte Carlo SSA}
\label{sec:MCSSA}

\subsection{Singular Spectrum Analysis}
Let us shortly describe the scheme of SSA to introduce notation used further
(see e.g. \cite{Golyandina.etal2018} for details).

Denote by $\tX = (x_1,\ldots,x_N)$ a time series of length $N$ and by $L$, $1<L<N$,
a window length.
The \emph{trajectory matrix} $\bfX = \cT(\tX)$ is determined as $\bfX = [X_1:\ldots:X_K]$,
where $K=N-L+1$ and $X_i = (x_i,\ldots,x_{i+L-1})^\rmT\in \spaceR^L$ are lagged vectors.

The next step is the SVD expansion $\bfX = \sum_{i=1}^d\sqrt{\lambda_i}U_i V_i^\rmT = \sum_{i=1}^d U_i (\bfX^\rmT U_i)^\rmT$, where
$\{U_i\}_{i=1}^d$ is the orthonormal system of the eigenvectors of the matrix $\bfX \bfX^\rmT$,
$\lambda_1\ge\lambda_2\ge\ldots\ge\lambda_d>0$ are the corresponding non-zero eigenvalues.
This is Basic SSA (or BK version); sometimes Toeplitz SSA (VG version) designed for stationary time series is considered.

The components of the obtained matrix decomposition are reasonably grouped and
each grouped matrix is transferred to a time series. Thus, the result of SSA is
a time series decomposition.

Although SSA is a model-free method, there is a model that fits it.
This is a class of signals, which are approximated by a sum of products
of polynomials, exponentials and sine waves.
In particular, a sum of sine waves is perfect for SSA. SSA
can extract sine waves with different frequencies if these frequencies are not too close
and can separate a sum of sinusoids from noise. If a sine wave series component can be extracted from
the time series (we say that it is separated from the residual by SSA), then the SVD decomposition
contains two eigenvectors $U_i$ and $U_{i+1}$, which have the same dominant frequency
as the original sine wave.

By the properties of the SVD, $\lambda_i = \|\Pi_{\{U_i\}} \bfX\|^2 = \|\bfX^\rmT U_i\|^2 =
\sum_{j=1}^K (X_j^\rmT U_i)^2$ can be interpreted as
the total squared norm of projections of lagged vectors to $\sspan(U_i)$.

\subsection{General comments}
On the one hand, Monte Carlo SSA is a well-developed method. In \cite{Allen.Smith1996}, the foundation of the method is thoroughly described. On the other hand, many questions are still under investigation; they are, among others, the best way of estimating the noise parameters and taking into consideration the presence of the nuisance signal. In this paper, we consider a most simple version of Monte Carlo SSA to concentrate on the problem of multiple testing.

In some cases, we will assume that the noise parameters are known. Also, we will show how can the criterion errors change if the parameters are estimated.

There is a natural modification of Monte Carlo SSA, which does not use the SSA decomposition step except for the construction of the trajectory matrix (a part of the study will be performed for such a modification). However, the relation with SSA is essential, since the detected signal can be reconstructed through SSA if we keep the connection with the eigenvectors in SSA.

\subsection{Single test}
\label{sec:single}
Let $\bm{\xi}=(\xi_1,\ldots,\xi_{N})$ be a red noise with parameters $\varphi$ and $\delta$.
Denote by $L$ the window length and by ${\bm\Xi}=\cT(\bm\xi)$ the trajectory matrix of $\bm\xi$.
Let a vector $\projVec\in \spaceR^L$, $\|\projVec\|=1$, be given. If we are interesting in a frequency contribution,
then $\projVec$ can be a sine wave with a given frequency.
The total squared norm of the projection of the columns of ${\bm\Xi}$ to the vector $\projVec$
is calculated as
\begin{equation*}
    p=\|{\bm\Xi}^\textrm{T}\projVec\|^2 = \projVec^\rmT \left({\bm\Xi}{\bm\Xi}^\textrm{T}\right)\projVec.
\end{equation*}

The null hypothesis states that the observed time series $\tX$ is a realization of $\bm\xi$ with some
parameters $\varphi$ and $\delta$.
Denote $\widehat{p}=\|{\bfX}^\textrm{T}\projVec\|^2$.
If $\projVec$ is an eigenvector of $\bfX\bfX^\rmT$, then $\widehat{p}$ is the corresponding eigenvalue.
Note that for a sinusoidal $\projVec$, $\widehat{p}$ negligibly depends on the phase of this sinusoid, since
  for large $K=N-L+1$ the lagged vectors consist of many shifts.

Let $\varphi$ and $\delta$ be known. Under some assumptions, the distribution of $p$ can be calculated theoretically.
Then the prediction interval with confidence level $\gamma$ is calculated as
the interval between $(1-\gamma)/2$- and $(1+\gamma)/2$- quantiles for the two-tailed test and
between zero and $\gamma$-quantile for the one-tailed test (upper-tailed test).
In both cases, $\widehat{p}$ belongs to the constructed predicted interval with probability $\gamma$.

If the theoretical distribution is unknown, then these quantiles can be calculated by simulation of
$G$ samples $\bm\xi^i$ of the random vector $\bm\xi$ and the use of empirical (sample) quantiles for the obtained sample $p_i=\|{\bm\Xi}_{i}^\textrm{T}\projVec\|^2$, $i=1,\ldots,G$.
The probability that $\widehat{p}$ belongs to the empirical (Monte Carlo) prediction interval tends to $\gamma$
as $G$ tends to infinity.

Recall that the significance level $\alpha$ is equal to $1-\gamma$ and therefore one can say that
the probability of the type I error $\alpha_I$ tends to $\alpha$.

For both theoretical considerations and simulations, the values of the parameters $\varphi$ and $\delta$ are used.
Here we do not discuss the estimation of noise parameters in the presence of a signal.
 Note that the proposed approach to hypothesis testing is valid for various modifications of the MC-SSA technique.

\begin{remark}
\label{rem:wrong}
It is important to note that $W$ may be produced by the time series itself. However, as was discussed in \cite{Allen.Smith1996}, then the surrogate data should be projected also to the vectors produced by them. There is a temptation to consider the version when the projection vector is produced by the observed time series and the test is constructed by projection to this vector. However, then the type-I error is not controlled and the test is liberal.
\end{remark}

\subsection{Choice of vectors for projection}
\label{sec:choice}
In practice, we do not know the frequency of a possible periodic signal component.
Therefore, the approach is to consider many vectors for projection,
which correspond to a set of frequencies.
For example, one can take a set of sine waves $W_1,\ldots, W_H\in \spaceR^L$ with equidistant frequencies from some
frequency interval $[\omega_1,\omega_2]\subset (0,0.5)$.
To obtain slightly dependent tests, the number of vectors should not exceed their dimension $L$.

The other choice is to take the set of eigenvectors produced by SSA (this is a common case in MC-SSA) with consideration to Remark~\ref{rem:wrong}.

The compromised version suggested in \cite{Allen.Smith1996} is $W_1,\ldots, W_H$ to be eigenvectors of the general-population covariance matrix. That is, the sample covariance matrix is $\bfX\bfX^\rmT$, where $\bfX$ is the $L$-trajectory matrix, whereas the general-population correlation matrix of red noise has the $(i,j)$-th term $\phi^{|i-j|}$. They are close; however, the difference from the viewpoint of type-I error can be drastic. If one wants to include to the set of projection vectors the sine wave vector with a specific frequency $\omega$, then the eigenvectors of the matrix with the $(i,j)$-th term $\phi^{|i-j|} + C\cos(2\pi \omega |i-j|)$ can be considered.

Thus, we will consider two choices:\\
1. $W_1,\ldots, W_H$ are the eigenvectors of the matrix
\begin{equation}
\label{eq:covar}
\{\phi^{|i-j|} + C\cos(2\pi \omega |i-j|)\}_{i,j=1}^L.
\end{equation}
2. $W_1,\ldots, W_H$ are the eigenvectors of the matrix $\bfX\bfX^\rmT$.

Since we study the version of Monte Carlo SSA, where the projections of surrogate data are performed on the fixed vectors, the second version is generally wrong. We will consider an example to show this.

\subsection{Multiple testing}
In Monte Carlo SSA, the prediction intervals are constructed for the contribution of each projection vector
independently.
Let $\projVec_1,\ldots,\projVec_H$ be a set of projection vectors. Denote $$\widehat{p}_k=\|\bfX^\textrm{T}\projVec_k\|^2, \qquad k=1,\ldots,H.$$
For each vector $\projVec_k$, the sample of squared
projection norms is constructed: $P_k=(p_{k1},\ldots,p_{kG})^\textrm{T}$, where $p_{ki}$ is calculated as
\begin{equation}
\label{eq:sample}
    p_{ki}=\|{\bm\Xi}_{i}^\textrm{T}\projVec_k\|^2, \qquad i=1,\ldots,G;
\end{equation}
here ${\bm\Xi}_{i}=\cT(\bm\xi^{i})$ is the trajectory matrix of the $i$th sample of red noise $\bm\xi^{i}=(\xi^{i}_1,\ldots,\xi^{i}_{N})$ with given parameters.

We can construct single prediction intervals for the contribution of each vector $W_k$ as it is described in
Section~\ref{sec:single}. The problem of multiple testing
(the problem of FWER, which can be much larger than the given significance level $\alpha$) can be solved by means of the Bonferroni correction. However, if the Bonferroni correction is used, the test becomes conservative (FWER is less than $\alpha$).
 To obtain an exact test, an approach similar to that of Tukey's HSD applied
to post hoc comparisons in ANOVA can be considered. That is, we can construct a test, which is based on the
distribution of the maximum of the standardized contributions $p_k$.
If this test rejects the null hypothesis, then all frequencies, which correspond to the projection vectors with the contribution lying outside the corrected prediction intervals,
are considered significant. Thus, we can talk about the significance of a frequency if put into correspondence projection vectors and frequencies. Note that for this approach, FWER is equal to $\alpha$.

Let us describe the algorithm of constructing the prediction intervals with correction for multiple testing.

The first version is straightforward; this is the single test for each vector $W_k$ with Bonferroni correction, that is, the significance level $\alpha/H$ is taken instead of $\alpha$.

\bigskip
   \textbf{Algorithm 1} (Single one-tailed test with Bonferroni correction)

   1. For each $k$, $k=1,\ldots,H$, calculate the test statistic $\hat{p}_k$ and the sample $P_k=\{p_{ki}\}_{i=1}^G$, see \eqref{eq:sample}.

   2. Find $q_k$ as the sample $(1-\alpha/H)$-quantile of $P_k$.

   3. The null hypothesis, which states that the time series is pure red noise, is not rejected if for each $k$  the inequality $\widehat{p}_{k}<q_k$ is valid; otherwise, the null hypothesis is rejected and a signal is detected.

   4. If $H_0$ is rejected, then post-hoc testing can be performed: the contribution of $W_k$ (and of the corresponding frequency) is significant if $\widehat{p}_k$ exceeds $q_k$. Thus, $[0, q_k]$ are considered as the corrected prediction intervals, $k=1,\ldots,H$.

\bigskip
   \textbf{Algorithm 2} (Single two-tailed test with Bonferroni correction)

   1. For each $k$, $k=1,\ldots,H$, calculate the test statistic $\hat{p}_k$ and the sample $P_k=\{p_{ki}\}_{i=1}^G$, see \eqref{eq:sample}.

   2. Find $q_k^{\text{low}}$ and $q_k^{\text{up}}$ as the sample $(0.5\alpha/H)$- and $(1-0.5\alpha/H)$-quantiles correspondingly.

   3. The null hypothesis, which states that the time series is pure red noise, is not rejected if for each $k$  $q_k^{\text{low}}<\widehat{p}_{k}<q_k^{\text{up}}$; otherwise, the null hypothesis is rejected and a signal is detected.

   4. If $H_0$ is rejected, then post-hoc testing can be performed: the contribution of $W_k$ (and of the corresponding frequency)
    is significant if $\widehat{p}_k$ does not belong $[q_k^{\text{low}},q_k^{\text{up}}]$. Thus, $[q_k^{\text{low}},q_k^{\text{up}}]$ are considered as the corrected prediction intervals, $k=1,\ldots,H$.

\medskip
Let us extend the multiple-testing approach, which was considered in \cite{Boyarov2012} in the framework of Monte Carlo SSA. The approach is based on the distribution of $\max_{1\leq k\leq H}(p_{ki}-\mu_k)/\sigma_k$, where $\mu_k$ and $\sigma_k$ are mean and standard deviation of $P_k$, $k=1,\ldots,H$. Here $\sigma_k$ reflects the size of the $k$-th prediction intervals. Normalization by $\sigma_k$ keeps the same difference between the intervals sizes as in single tests. By construction, the size of a prediction interval is related to the test power against the existence of the signal at the corresponding  frequency. Therefore, we can consider $\widetilde\sigma_k = w_k\sigma_k$, where $w_k$ is a weight; an increase of the weight corresponds to more expectation of a signal at the frequency matched to $W_k$. Some weights $w_k$ can be zero; this means that the corresponding projection vectors do not participate in the testing and the test is disabled to detect the signal at the corresponding frequencies. Thus, it is the same, to set zero weights outside the frequency range (and unit weights within it) or just consider the subset of the projection vectors matched to the frequency interval. However, the approach with weights is much more flexible and allows to take arbitrary weights between 0 and 1.

\bigskip
   \textbf{Algorithm 3} (Weighted multiple one-tailed test)

   1. For each $k$, $k=1,\ldots,H$, calculate the test statistic $\hat{p}_k$, the sample $P_k=\{p_{ki}\}_{i=1}^G$, see \eqref{eq:sample}, and calculate its mean $\mu_k$ and standard deviation $\sigma_k$.

   2. Calculate $\bm{\eta}=(\eta_1,\ldots,\eta_G)$, where \[\eta_i=\max_{1\leq k\leq H}(p_{ki}-\mu_k)/\widetilde\sigma_k,\ \widetilde\sigma_k = w_k\sigma_k, \qquad i=1,\ldots,G.\]

   3. Find $q$ as the sample $(1-\alpha)$-quantile of $\bm{\eta}$.

   4. The null hypothesis, which states that the time series is pure red noise, is not rejected if \[\max_{1\leq k\leq H}(\widehat{p}_{k}-\mu_k)/\widetilde\sigma_k<q;\] otherwise, the null hypothesis is rejected and a signal is detected.

   5. If $H_0$ is rejected, then post-hoc testing can be performed: the contribution of $W_k$ (and of the corresponding frequency) is significant if $\widehat{p}_k$ exceeds $\mu_k + q w_k\sigma_k$. Thus, $[0,\mu_k + q w_k \sigma_k]$ are considered as the corrected prediction intervals, $k=1,\ldots,H$.

\bigskip
   \textbf{Algorithm 4} (Weighted multiple two-tailed test)

   1. For each $k$, $k=1,\ldots,H$, calculate the test statistic $\hat{p}_k$, the sample $P_k=\{p_{ki}\}_{i=1}^G$, see \eqref{eq:sample}, and calculate its mean $\mu_k$ and standard deviation $\sigma_k$.

   2. Calculate $\bm{\eta}=(\eta_1,\ldots,\eta_G)$, where \[\eta_i=\max_{1\leq k\leq H}|p_{ki}-\mu_k|/\widetilde\sigma_k,\ \widetilde\sigma_k = w_k\sigma_k,\qquad i=1,\ldots,G.\]

   3. Find $q$ as the sample $(1-\alpha)$-quantile of $\bm{\eta}$.

   4. The null hypothesis, which states that the time series is pure red noise, is not rejected if \[\max_{1\leq k\leq H}|\widehat{p}_{k}-\mu_k|/\widetilde\sigma_k<q; \] otherwise, the null hypothesis is rejected and a signal is detected.

   5. If $H_0$ is rejected, then post-hoc testing can be performed for a fixed $k$: the contribution of $W_k$ (and of the corresponding frequency) is significant if $|\widehat{p}_k - \mu_k|/\widetilde\sigma_k$ exceeds $q$. Thus, $[\mu_k - q w_k\sigma_k,\mu_k + q w_k\sigma_k]$ are considered as the corrected prediction intervals, $k=1,\ldots,H$.

\medskip
Algorithms 2 and 4 correspond to two-sided prediction intervals (this is the conventional version of Monte Carlo SSA), while Algorithms 1 and 3 correspond to the one-sided case.  This corresponds to hyper-rectangular and half
hyper-rectangular prediction regions respectively.

   \medskip
   \begin{remark}
   The choice of the vectors $W_k$, $k=1,\ldots,H$, was discussed in Section~\ref{sec:choice}. To increase the criterion power, the number $H$ of the vectors should be as small as possible, e.g., only vectors with dominant frequencies from a given frequency range can be taken. If the vectors $W_k$ are sine waves, the choice of vectors with frequencies from the given range is trivial. If $W_k$ are the eigenvectors, then their dominant frequencies can be calculated by e.g. the ESPRIT method \cite[Section 3.1]{Golyandina.etal2018}. This has a little sense for white noise, for which each eigenvector is a mixture of a lot of frequencies. However, for red noise, it has sense, since the eigenvectors correspond to narrow ranges of frequencies. As we mentioned above, increasing the power can be performed by setting the weights.
   \end{remark}

\section{Numerical investigation}
\label{sec:study}
    Let us introduce abbreviations for the test versions. We consider 3-symbols abbreviations with an optional information
    \begin{equation}
    \label{abbrev}
    \text{\{M,S,B\}\{T,E\}\{1,2\}[Est]}.
    \end{equation}
    The first letter M (Multiple) corresponds to Algorithms 3 and 4, S (Single) corresponds to Algorithms 1 and 2 without Bonferroni correction, while B means using this correction. The second letter is related to the way of generating the projection vectors: T means that the eigenvectors of the theoretical covariance matrix \eqref{eq:covar} of red noise ($C=0$) are considered; E means that the eigenvectors of the empirical covariance matrix $\bfX\bfX^\rmT$ are used. The last digit means what (one- or two-tailed) test is considered.  We consider the method MT1 as the basic one, which is exact and more powerful.

    This abbreviation corresponds to the use of true parameters of red noise and unit weights $w_k$. If the noise parameters are estimated, then we will add `Est' after the digit. If another modification is used, this is indicated separately.

    \R{}-scripts in \cite{Golyandina2021} contain an implementation of the above algorithms.

Let us demonstrate the results of different versions of Monte Carlo SSA.

\subsection{Example}
The model of an artificial time series is
\begin{equation}
\label{eq:ex_sin}
x_n = A \sin(2\pi \omega n) + \xi_n,
\end{equation}
 where $\xi_n$ is red noise with parameters $\varphi$ and $\delta$, $n=1,\ldots,N$.  The case $A=0$ corresponds to the null hypothesis and the case $A>0$ yields the presence of signal, that is, corresponds to an alternative. Hereafter we consider the  AR(1) parameters $\varphi = 0.7$ and $\delta = 1$.

For illustrative examples, we take $N=1000$ and the signal parameters $A=0.4$ and $\omega=0.2$ in $H_1$.
The parameters of MC-SSA are $L=40$ and $G=1000$.

Let us consider a fixed significance level 0.2 (that is, the confidence level equals 0.8).
   To weaken the dependence on the time series length $N$, we consider $\|\tX\|^2 = \sum_{i=1}^N x_i^2/N$.
   We used the true parameters of AR(1) for the creation of surrogate data. The continuous curve
   is the spectral density of AR(1) with the parameters that were used in the simulation.
   We calculated the dominant frequencies of $\projVec_k$ by the ESPRIT method with the rank $r=2$.

\begin{figure*}[!ht]
        \begin{center}
        \includegraphics[width=0.45\textwidth]{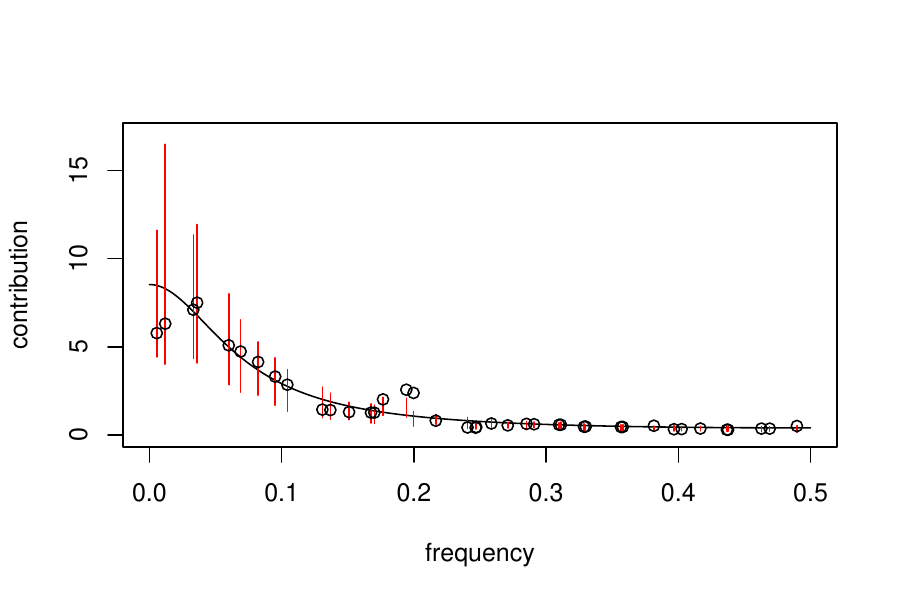}
        \includegraphics[width=0.45\textwidth]{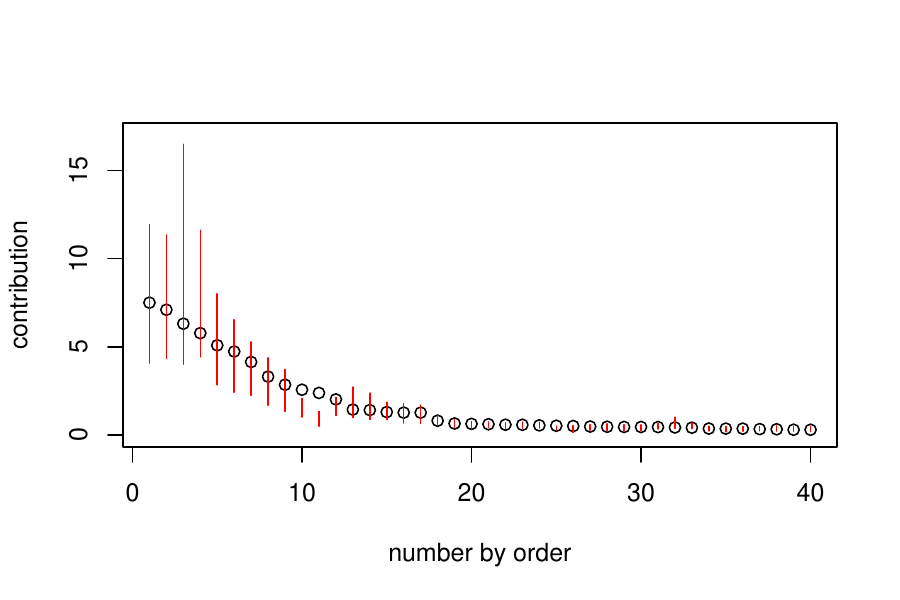}
        \end{center}
        \caption{ME2.
        Left --- ordered by frequency, right --- ordered by projection value ($\hat{p}_k$).}
        \label{fig:multi}
\end{figure*}

\begin{figure*}[!ht]
        \begin{center}
        \includegraphics[width=0.45\textwidth]{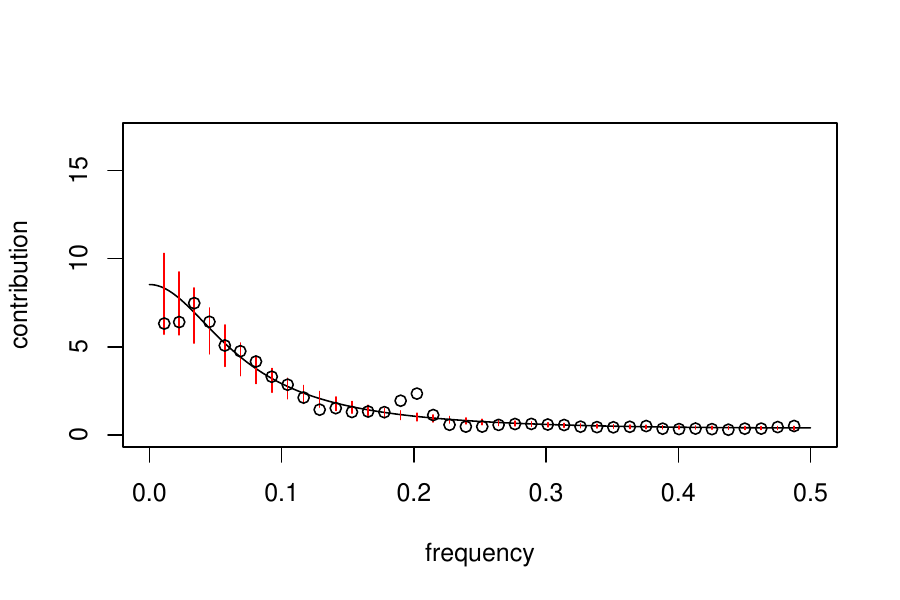}
        \includegraphics[width=0.45\textwidth]{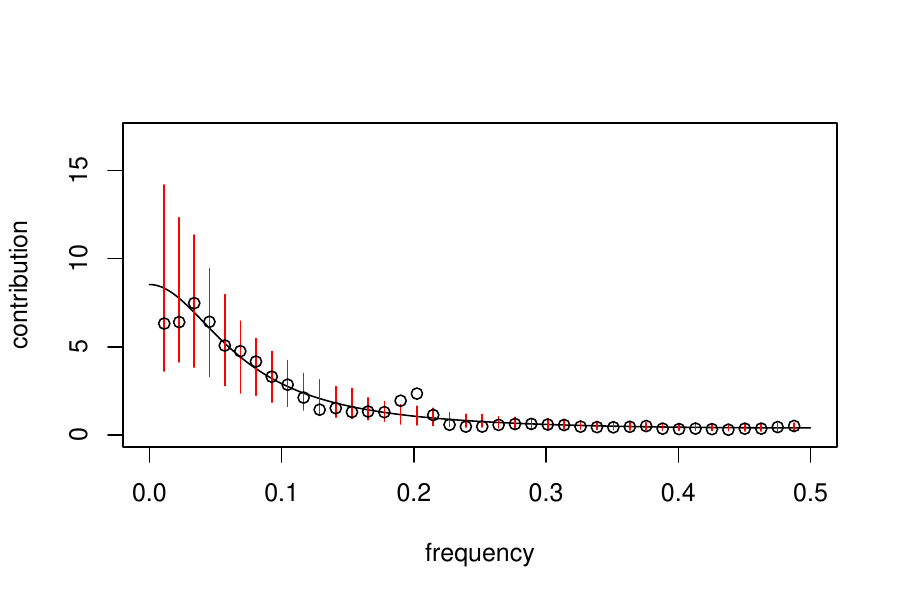}
        \end{center}
        \caption{\{S,B\}T2.
        Left --- without correction, right --- with Bonferroni correction.}
        \label{fig:single}
\end{figure*}

\begin{figure*}[!ht]
        \begin{center}
        \includegraphics[width=0.45\textwidth]{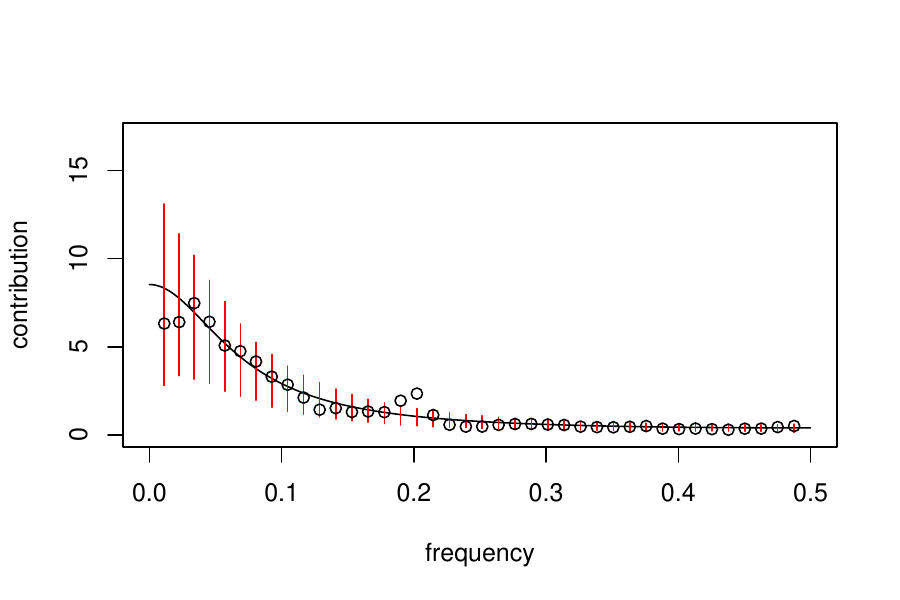}
        \includegraphics[width=0.45\textwidth]{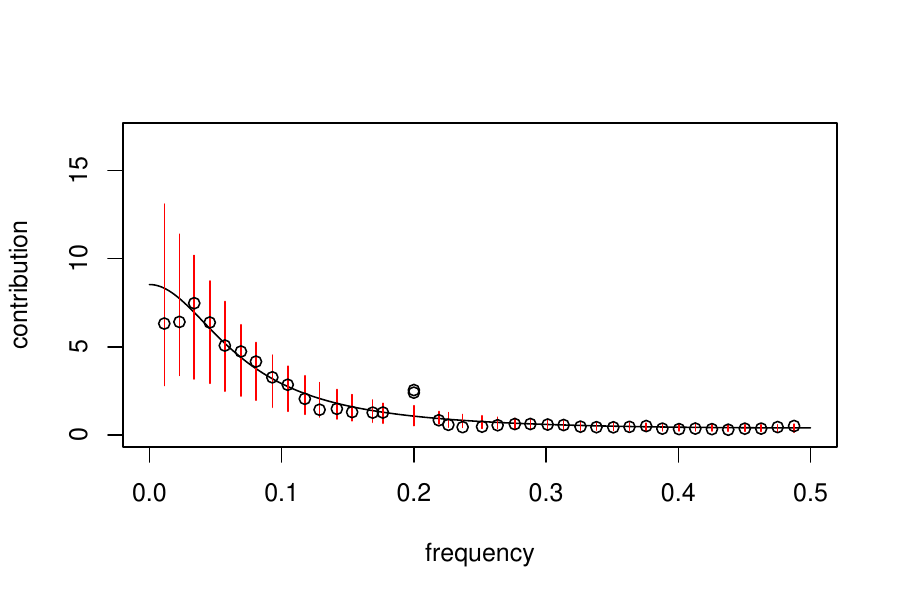}
        \end{center}
        \caption{MT2.
        Left --- eigenvectors of pure red-noise covariance matrix, right ---  eigenvectors of the covariance matrix of red-noise with an added sinusoid at frequency 0.2.}
        \label{fig:tmulti}
\end{figure*}

\begin{figure*}[!ht]
        \begin{center}
        \includegraphics[width=0.45\textwidth]{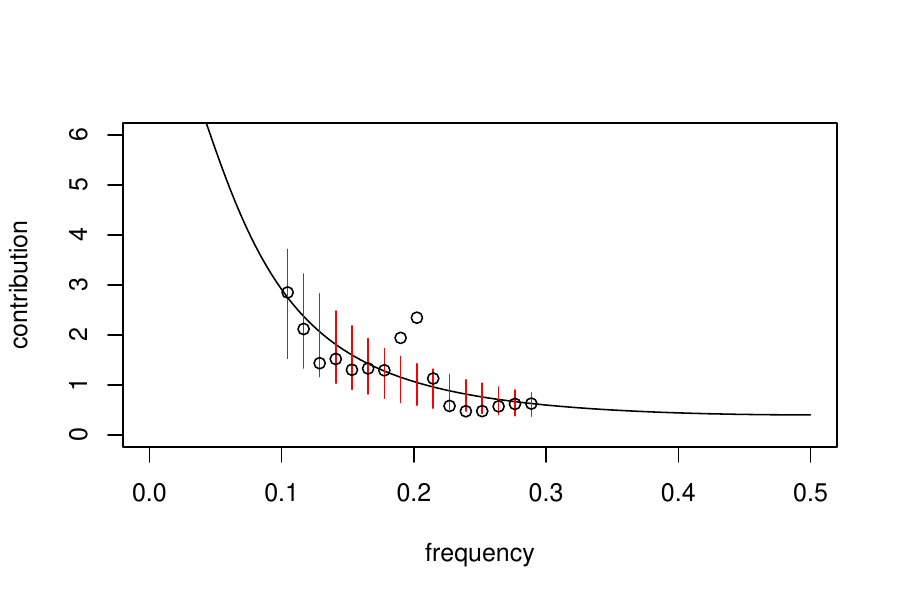}
        \includegraphics[width=0.45\textwidth]{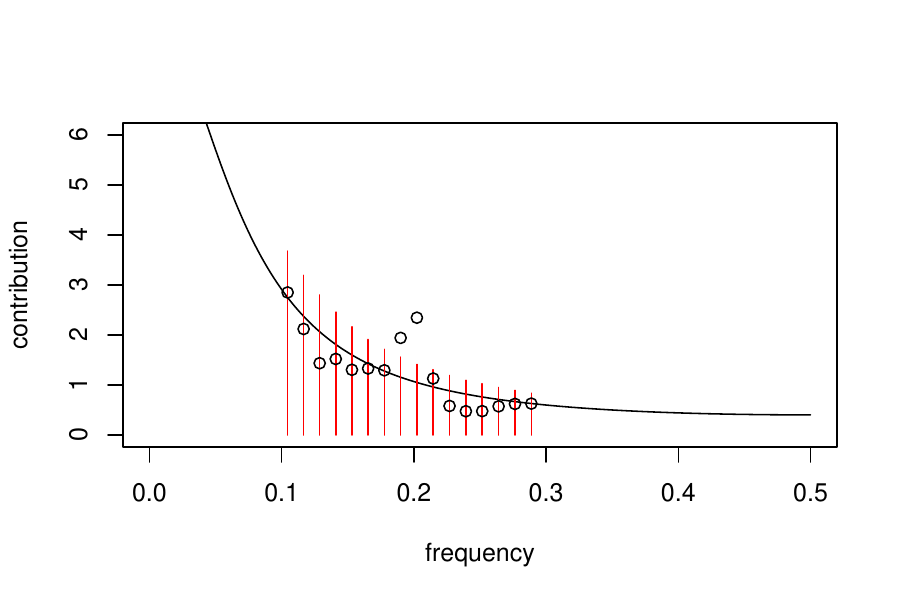}
        \end{center}
        \caption{MT\{1,2\},
        frequency range $[0.1,0.3]$. Left --- two-tailed test, right --- one-tailed test.}
        \label{fig:tmulti_range}
\end{figure*}

\begin{figure*}[!ht]
        \begin{center}
        \includegraphics[width=0.45\textwidth]{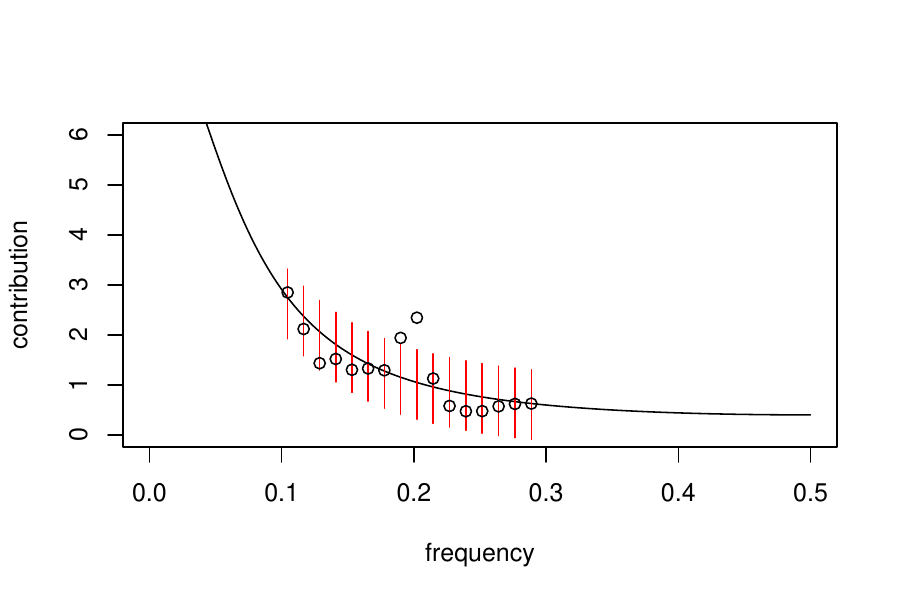}
        \includegraphics[width=0.45\textwidth]{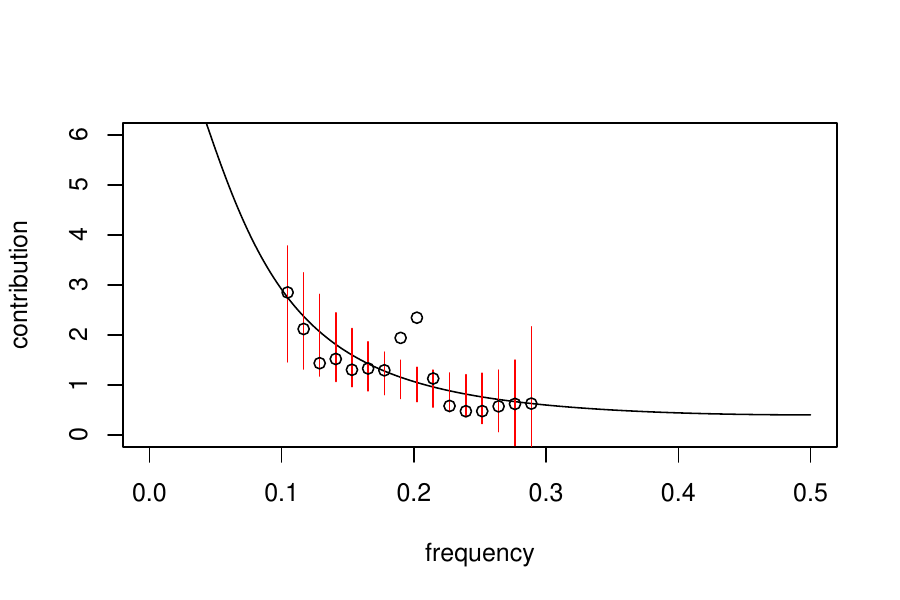}
        \end{center}
        \caption{MT2, weighted,
        frequency range $[0.1,0.3]$,  two-tailed test. Left --- the weights to obtain equal prediction intervals ($w_k = 1/\sigma_k$), right --- the weights to increase the power against frequencies around 0.2.}
        \label{fig:tmulti_range_weighted}
\end{figure*}

Although one-tailed versions of the tests are more powerful, on most of the figures we show two-sided prediction intervals for  a clearer visual presentation.

The presence of a sine-wave signal mostly corresponds to exceeding the upper bound. Therefore, as a rule, for exact tests, if the upper bound is smaller, then the test is more powerful.

Fig. \ref{fig:multi} demonstrates generally a liberal test, which therefore cannot be used in practice as is. In this version the projection vectors are generated by the observed time series; this choice corresponds to a popular version of Monte Carlo SSA.

Fig. \ref{fig:single} shows prediction intervals for the single test with projection vectors chosen as theoretical eigenvectors; the Bonferroni correction is used in Fig.~\ref{fig:single} (right). The test in Fig. \ref{fig:single} (left) is strongly liberal, while Fig.~\ref{fig:single} (right) is slightly conservative.

Fig. \ref{fig:tmulti} contains multiple prediction intervals, which provide the exact test with theoretical eigenvectors as the projection vectors.
In Fig. \ref{fig:tmulti} (left), the eigenvectors of the matrix \eqref{eq:covar} with $C=0$ are taken as $W_k$. The upper bounds are slightly lower than in Fig. \ref{fig:single} (left).
In Fig. \ref{fig:tmulti} (right), the information about the presence of a signal of frequency $\omega = 0.2$ is used and the eigenvectors of the matrix \eqref{eq:covar} with $C=1$ are taken as $W_k$.

Figs. \ref{fig:tmulti_range} and \ref{fig:tmulti_range_weighted} are related to the case of zero weights for the frequencies of the projection vectors outside the frequency interval $[0.1,0.3]$.
Fig. \ref{fig:tmulti_range} demonstrates the difference between two-sided and one-sided prediction intervals. In the one-sided case, the upper bound is slightly lower.
Fig. \ref{fig:tmulti_range_weighted} shows how the weights influence the prediction intervals sizes. We set \code{weights = c(seq(6.25,8,0.25), 8:1)}, what means that the weights increase from 6.25 to 8 with step 0.25 and then decrease from 8 to 1. Influencing weights on the test power will be studied later.

\subsection{Study of statistical properties of MC-SSA}
Let us describe the methodology of the study of statistical properties of the
constructed criteria with bootstrapping. 
The key items are:

\begin{enumerate}
\item
The first step is to simulate synthetic data according to the null hypothesis $M$ times and estimate $\alpha_I = \alpha_I(\alpha)$ as the proportion of the rejected null hypothesis for a given significant level $\alpha$. Check that the necessary condition $\alpha_I\le \alpha$ is fulfilled. It is necessary to find the sufficient size $G$ of the surrogate data.
\item
If the necessary condition is fulfilled, then the second step is to simulate synthetic data according to an alternative hypothesis $M$ times and estimate the power $1 - \alpha_{II}$ against this hypothesis as the proportion of the rejected null hypothesis for a given significant level $\alpha$.
\item
Compare different criteria by power against the alternative hypothesis under interest and use the one with the larger power.
\item
If there is no test with sufficient power, then it is possible to improve a test, where $\alpha_I< \alpha$ or $\alpha_I> \alpha$.
Then find $\widetilde\alpha$ such
that $\alpha_I(\widetilde\alpha) \approx \alpha$ and
use the significance level $\widetilde\alpha$ instead of
$\alpha$.
\end{enumerate}

Below we consider these items.
As before, we consider the  AR(1) parameters $\varphi = 0.7$ and $\delta = 1$.
In a case, when the AR(1) parameters are estimated, the maximum likelihood method was applied, where the conditional-sum-of-squares method was used to find starting values.

\medskip
\textbf{Choice of $G$.}
A sufficient value $G$  in item 1 can be determined theoretically. The value of $G$ should be considerably large to estimate the quantiles for the surrogate data.  In the case of single prediction intervals with Bonferroni correction, $(1-\alpha/H)$-quantiles should be estimated (let us consider one-tailed tests), while in the case of multiple testing, we need $(1-\alpha)$-quantiles only. For example, if $\alpha = 0.05$, for multiple testing, $G = 1000$ is enough, since the estimate is the 50-th value from the maximum in an ordered sample. However, for single testing and $H=100$ projection vectors, $\alpha/100 = 0.005$ and therefore the 0.995-quantile will be underestimated (and therefore the test will be liberal).
Table~\ref{tab:powerG} contains the estimates of the type I errors for $N=L=200$ and $\alpha=0.2$ (then the Bonferroni correction gives the significance level $0.001$); $M=10000$ simulations were used. Table~\ref{tab:powerG} confirms that the multiple methods need smaller $G$. Hereinafter, the columns `2.5\%' and `97.5\%' show the bounds of 95\%-confidence intervals for the estimated probabilities (the type I error or the power).

\begin{table}[!ht]
\centering
\caption{Type-I error for different $G$.}
\begin{tabular}{|l|r|r|r|r|}
  \hline
   Method & $G$ & est. type I error & 2.5\% & 97.5\% \\
  \hline
  MT1& 100&0.466 & 0.456 & 0.476 \\
  BT1& 100&0.864 & 0.857 & 0.870 \\
  \hline
  MT1& 500&0.250 & 0.241 & 0.258 \\
  BT1& 500&0.326 & 0.316 & 0.335 \\
  \hline
  MT1& 1000&0.219 & 0.211 & 0.227 \\
  BT1& 1000&0.212 & 0.204 & 0.221 \\
  \hline
\end{tabular}
\label{tab:powerG}
\end{table}

\begin{figure*}[!ht]
        \begin{center}
        \includegraphics[width=0.45\textwidth]{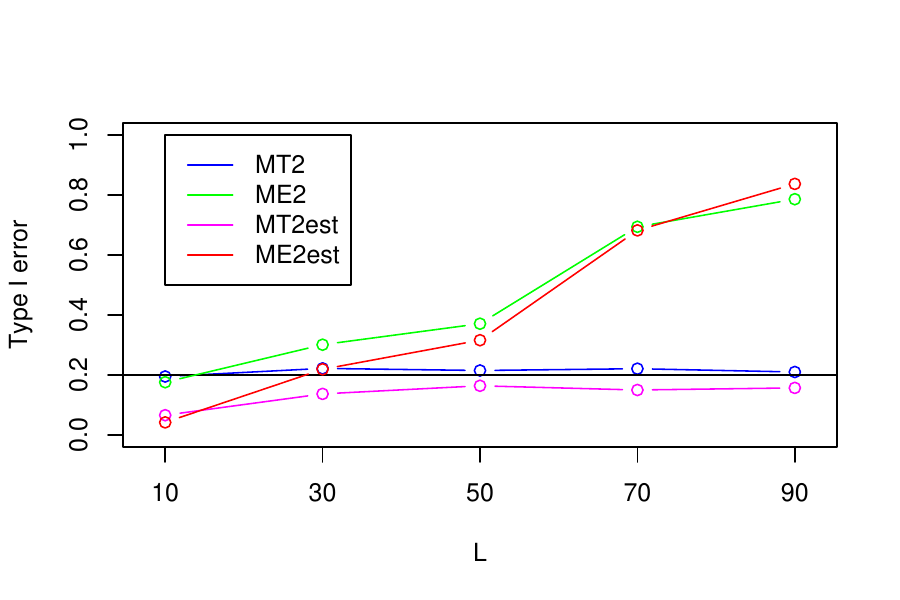}
        \includegraphics[width=0.45\textwidth]{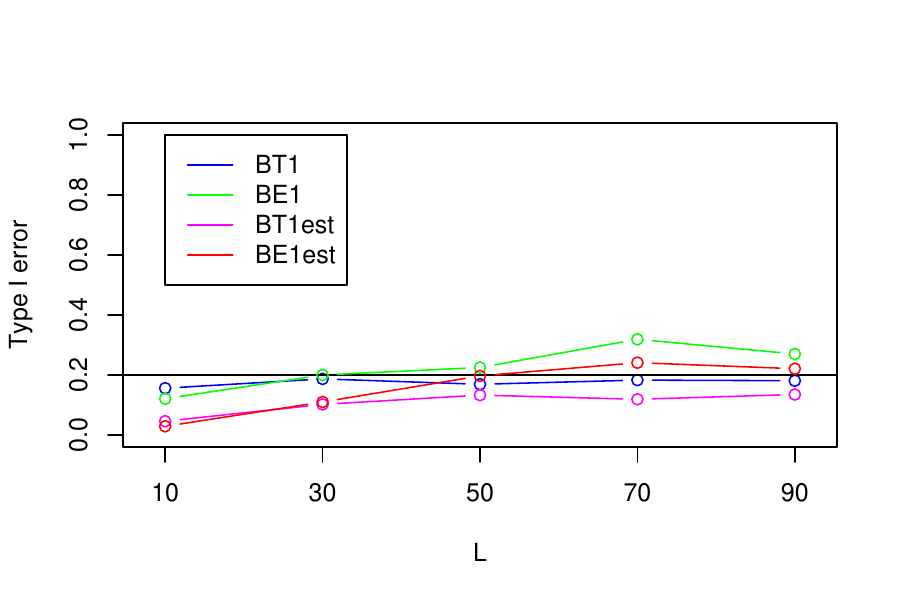}
        \end{center}
        \caption{Type I error for different $L$, empirical and theoretical eigenvectors. Left --- the multiple test, right --- the single test with the Bonferroni correction.}
        \label{fig:ev_t}
\end{figure*}

\medskip
\textbf{Type I error.}
The estimates of the family-wise type I errors  are contained in Fig.~\ref{fig:ev_t};
we use $N = 100$, $L=10$, $M=1000$, and $G=1000$.

\smallskip
Fig.~\ref{fig:ev_t} (left) demonstrates the difference between the use of theoretical and empirical eigenvectors. The pluses and minuses of these versions are as follows.

1. Multiple test, empirical eigenvectors.

Plus: It is possible to reconstruct the signal on the basis of significant projection vectors (empirical eigenvectors).

Minus: The test is liberal.

2. Multiple test, theoretical eigenvectors.

Plus: The test is exact.

Minus: If the SSA reconstruction of a detected signal should be performed, a paring is needed to put into correspondence theoretical and empirical eigenvectors.

\smallskip
Fig.~\ref{fig:ev_t} (right) shows the difference between the use of multiple and single tests.
The pluses and minuses of these versions are as follows.

1. Multiple test

Plus: The test is exact.

Minus: Recalculation is necessary to consider a subset of the projection vectors.

2. Single test

Plus: It is easy to consider subsets of projection vectors without recalculation.

Minus: For the version without correction, it is a very liberal test.
For the version with Bonferroni correction, it is a slightly conservative test (this is visible for small $L$) and large $G$ is needed (visible for large $L$).

\medskip
In addition, Fig.~\ref{fig:ev_t} shows that if the noise parameters are estimated, then the MT and BT tests becomes very conservative (the type I error is
considerably smaller than the given level $\alpha = 0.2$). Also, the conservativeness of Bonferroni correction can compensate for the liberality of the test \{M,B\}E;  it can be a reason why the BE test is less liberal than the ME one in Fig.~\ref{fig:ev_t}.

\medskip
\textbf{Power.}
Let us estimate the power of different versions of the Monte Carlo SSA tests.
Consider the presence of a signal in $H_1$ with $A = 1$ and study the dependence of test power on the signal frequency. We used a standard method of estimation of AR(1) parameters, which ignores a possible presence of a signal.

\begin{figure}[!ht]
        \begin{center}
        \includegraphics[width=0.45\textwidth]{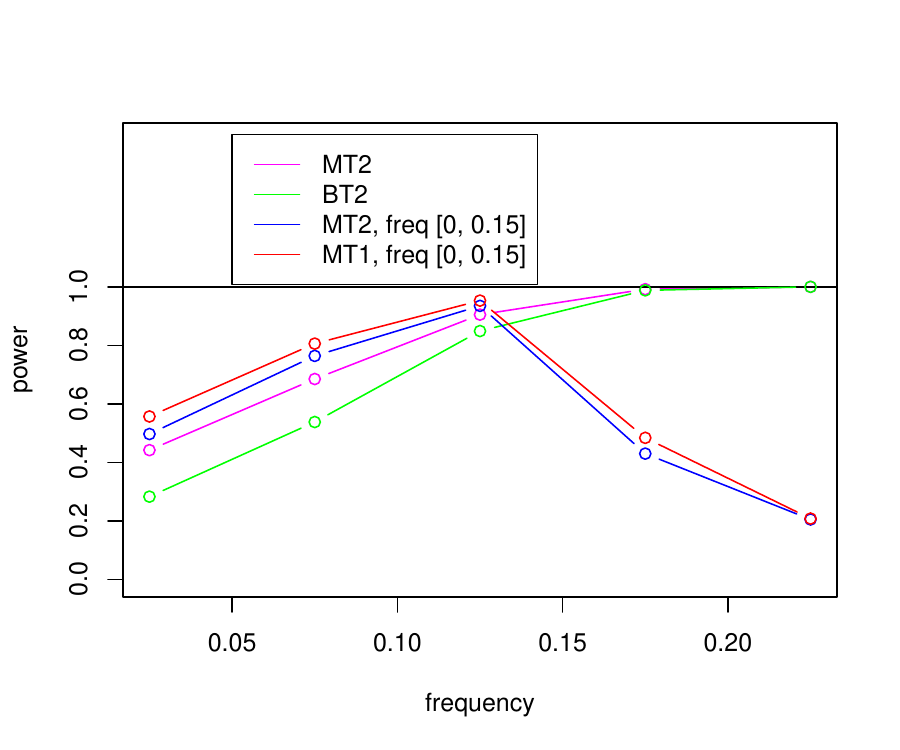}
        \end{center}
        \caption{Power for alternatives with sinusoidal signals of different frequencies.}
        \label{fig:power}
\end{figure}

The estimates of power are depicted in Fig.~\ref{fig:power}.
One can see that both the one-tailed version and a narrow frequency interval can increase the test power.

\medskip
\textbf{Correction of type I error.}
Let us illustrate item 4 of the described scheme with the help of the example \eqref{eq:ex_sin}, where $H_0$ corresponds to $A=0$ and $H_1$ corresponds to $A=1$, $\omega=0.1$.
As before, we consider a fixed significance level 0.2.

If the noise parameters are estimated, then the criterion becomes very conservative (the type I error is
two times smaller than the given $\alpha = 0.2$).
Changing $\alpha$, we can find that for $\widetilde\alpha=0.35$ we obtain $\alpha_I \approx 0.2 = \alpha$ (Table~\ref{tab:power}, the first three rows).

The criterion with the adjusted formal significance level is
more powerful than the original conservative criterion.
The last three rows of Table~\ref{tab:power} show that the use of the conservative test MT1est decreases the power in comparison with that of MT1, while after correction of the significance level to $\widetilde\alpha=0.35$, the power of MT1est considerably increases.

\begin{table}[!ht]
\centering
\caption{Improvement of the power of a conservative test.}
\begin{tabular}{|l|r|r|r|}
  \hline
   & est. type I error & 2.5\% & 97.5\% \\
  \hline
  MT1& 0.200  & 0.176 & 0.226 \\
  MT1est($\widetilde\alpha = 0.2$)& 0.069 & 0.054 & 0.086 \\
  MT1est($\widetilde\alpha = 0.35$)& 0.208  & 0.183 & 0.234 \\
  \hline
   & est. power & 2.5\% & 97.5\% \\
  \hline
  MT1& 0.800  & 0.774 & 0.824 \\
  MT1est($\widetilde\alpha = 0.2$)& 0.575 & 0.544 & 0.606 \\
  MT1est($\widetilde\alpha = 0.35$)& 0.731  & 0.702 & 0.758 \\
  \hline
\end{tabular}
\label{tab:power}
\end{table}

Note that the dependence of the test power on the significance level shows the test strength against a chosen alternative hypothesis. To find how the test would work if the formal significance level is corrected to obtain a given probability of type I error the dependence of the test power on the probability of type I error should be studied; this dependence is a kind of ROC curve.

\section{Conclusion}
In this paper, we studied Monte Carlo SSA from a statistical viewpoint.
A scheme for checking the MC-SSA algorithms for correct multiple type I errors and comparing the algorithms by power was proposed to avoid wrong conclusions in practice.  We demonstrated this scheme on a basic version of MC-SSA. However, it can be applied in a general case.

The numerical comparison of multiple and single versions (with Bonferroni correction), of different choices of the projection vectors, and of two-tailed and one-tailed tests was performed. Also, we considered weighted versions of the multiple MC-SSA.
Our recommendation is to use the multiple one-tailed MC-SSA; weights are applied if information about the possible frequencies is available. Also, we propose to correct the significance level if the used test is liberal or conservative.

\section*{Acknowledgment}
\addcontentsline{toc}{section}{Acknowledgment}
We are grateful to Alex Shlemov for the fruitful discussions.

\bibliographystyle{plain}

\end{document}